\documentclass[%
 aip,
 jmp,%
 amsmath,amssymb,
 preprint,%
]{revtex4-1}

\usepackage{amsmath}
\usepackage{textcomp}
\usepackage{epstopdf} 
\usepackage{amssymb}
\usepackage{wrapfig}
\usepackage{ragged2e}
\usepackage{graphicx}
\usepackage[T1]{fontenc}
\usepackage{lmodern}
\usepackage{bm}
\usepackage{natbib}
\usepackage{balance}  
\usepackage{times}    
\usepackage{url}      
\usepackage{soul}
\usepackage{color}
\usepackage{subfig}
\usepackage{amsmath,amsthm, amssymb, latexsym}

\usepackage{graphicx}
\usepackage{dcolumn}
\usepackage{bm}

\begin{document}

\preprint{}

\title[]{Modified Stoney's equation with anisotropic substrates undergoing large deformations}

\author{Sai Sharan Injeti}
\author{Nihit Vyas}%
\author{Sundararajan Natarajan}

\author{Ratna Kumar Annabattula}
\email{ratna@iitm.ac.in}

\date{\today}
\affiliation{%
Department of Mechanical Engineering, Indian Institute of Technology Madras, Chennai 600036, Tamil Nadu, India}
\begin{abstract}
	Residual stresses in a thin film deposited on a substrate results in a curvature of the system, which can be measured using the well known Stoney equation. Isotropic elasticity of the substrate along with infinitesimal strains and rotations are two important assumptions used in the derivation of the Stoney equation. However, the transverse deflection in the substrate contributes significantly to the extensional strain in its plane, leading to non-linearity in its deformation. Moreover, Silicon wafers are predominantly used as substrate materials to measure the curvature of the system. In this paper, relations between normalized substrate curvature and normalized thin film mismatch are derived in the non-linear deformation regime, for substrates made of single crystal Si(001) and Si(111) wafers. Numerical results of curvature of thin film configurations with Si(001) and Si(111) wafer substrates, undergoing large deformations are presented and discussed.
\end{abstract}

\maketitle

\section{\label{sec:level1}Introduction}
The thin film - substrate configuration assumes a curvature due to a mismatch strain (e.g., elastic, thermal) between the film and the substrate. The curvature of such systems can be expressed in terms of the residual stress present in the film through the Stoney equation\cite{Flinn1987}.
\begin{equation}
\sigma_f=\frac{E_s h_s^2}{6(1-\nu_s)h_f}\kappa,\label{eq:stoney-original}
\end{equation}
where $\sigma_f$ is the equi-biaxial residual stress in the film, $\kappa$ is the uniform spherical curvature of the film-substrate configuration, $h_f$ is the film thickness and $h_s$ is the thickness of the substrate. Eq.~\ref{eq:stoney-original} assumes the substrate material to be elastically isotropic with $E_s$ being its modulus of elasticity and $\nu_s$ being the Poisson's ratio. Several assumptions\cite{Freund1999, Freund2000} are made while deriving Eq.~\eqref{eq:stoney-original}. Assumptions that the strains and rotations are infinitesimally small and that the substrate material is elastically isotropic will be relaxed in this paper. Modifications to the original Stoney equation have been gaining importance since its first appearance. This is evident from papers published in recent years addressing or enhancing its accuracy using improved methods \cite{Claude2000, Zhang2005, Zhang2009} to extending it to a non-uniform stress state \cite{Blech2005}. Modified curvature relations were proposed by \citet{Nix1989}, which use Silicon wafer substrates. Later, the modified Stoney equation considering thin and elastically isotropic substrates was derived by \citet{Freund1999}. The equation that relates the substrate curvature to the thin film mismatch in the non-linear deformation range for elastically isotropic and thick substrates is given by \citet{Freund2000}
\begin{equation}
S=K[1+(1-\nu_s)K^2],\label{eq:freund-largedef} \quad S=\dfrac{3}{2}\dfrac{\epsilon_mR^2h_fE_f(1-\nu_s)}{h_s^3E_s(1-\nu_f)}, \quad K=\dfrac{1}{4}\dfrac{R^2\kappa}{h_s},
\end{equation}
where $S$ is the normalized mismatch strain in the film and $K$ is the normalized curvature. 
Stress-curvature relations for thick and anisotropic substrates in the small deformation regime were derived by~\citet{Janssen2008}. The stress-curvature relation for Si(001) wafers is given by
\begin{equation}
\sigma_fh_f=\frac{h_s^2}{6(s_{11}^{Si}+s_{12}^{Si})R}.\label{eq:stoney-si001}
\end{equation}
The stress-curvature relation for Si(111) wafers is given by
\begin{equation}
\sigma_fh_f=\left(\frac{6}{4s_{11}^{Si}+8s_{12}^{Si}+s_{44}^{Si}}\right)\frac{h_s^2}{6R},\label{eq:stoney-si111}
\end{equation}
where $s_{ij}^{Si}$ are elements of the compliance matrix of Si.\\
The anisotropy in material properties may lead to anisotropy in the stress state\cite{Zhao2002}. But the discussion in this paper is restricted to a system in which the film is under the influence of an equibiaxial stress. 

\subsection{\label{sec:level2}Scope of the paper}

In this paper, Eq.~\eqref{eq:freund-largedef} is extended to configurations with single crystal Si(001) and Si(111) wafer substrates. These equations are derived by minimizing the potential energy of the system. A similar analysis was used by \citet{Sharan2016} to derive the Stoney equation for systems with thin and anisotropic substrates in the small deformation regime. Numerical results for curvatures of Si wafer substrates in the non-linear deformation range are presented. Deviations of curvatures obtained from the derived equations and equations ~\eqref{eq:stoney-si001} and ~\eqref{eq:stoney-si111} with respect to numerical results are discussed, in a broad range of thin film mismatch.

\section{\label{sec:level1}Mathematical Formulation and Derivation}
In this section, a circular film-substrate system is analyzed with the assumption of uniform curvature for simple analytical treatment. However, it is to be noted that the curvature of the system in the non-linear deformation regime varies across the plane of the substrate \cite{Finot1997}. This variation is captured by the numerical results presented in the next section and is compared with the derived result. The stress distribution across the film thickness is assumed to be uniform. 
\begin{figure}[!h]
\begin{center}
\includegraphics[width=8cm]{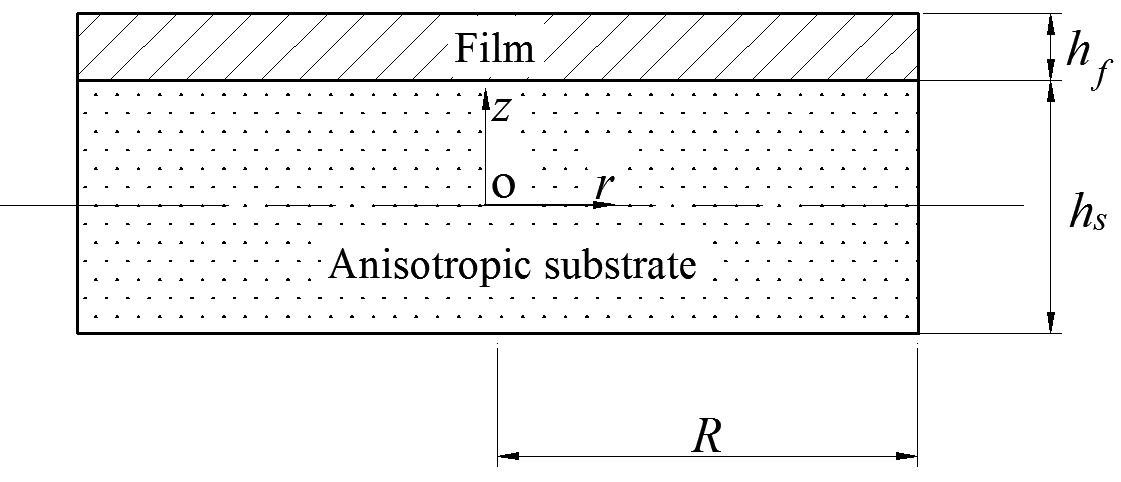}
\caption{Schematic of a circular thin film deposited on an anisotropic substrate}
\label{fig:Figure-1}
\end{center}
\end{figure}
Figure~\ref{fig:Figure-1} shows the cross sectional view of the thin film system. $h_s$, $h_f$ and $R$ represent the thickness of the substrate, thickness of the film and radius of the circular system, respectively. Deformation in the configuration is measured using a cylindrical coordinate system ($r$,$\theta$,$z$). The origin of the coordinate system is considered to be at the intersection of the substrate mid-plane and the axis of symmetry of the system. In this work, a radially symmetric deformation is considered due to the uniformity in film stress and the circular substrate geometry. The radial stress ($\sigma_{rr}$) and the circumferential stress ($\sigma_{\theta\theta}$) are the only non zero components of stress in the substrate and film because the deformation is axially symmetric and the out-of-plane stress ($\sigma_{zz}$) is assumed to be negligible. Hence, the elastic strain energy density in the film and substrate materials can be represented as
\begin{equation}
U(r,z)=\dfrac{1}{2}(\sigma_{rr}\epsilon_{rr}+\sigma_{\theta\theta}\epsilon_{\theta\theta}),\label{eq:energy-density-expanded}
\end{equation} 
where $\epsilon_{rr}$ and $\epsilon_{\theta\theta}$ are the radial and circumferential strain components, respectively. Let the radial and transverse displacements at a point on the substrate mid-plane be represented by $u(r)$ and $w(r)$, respectively. Then, $\epsilon_{rr}$ and $\epsilon_{\theta\theta}$ for large deformations can be written as\cite{Freund1999}
\begin{equation}
\epsilon_{rr}(r,z)=u'(r)-z w''(r)+\dfrac{1}{2}w'(r)^2+\epsilon_{m},\quad 
\epsilon_{\theta\theta}(r,z)=\dfrac{u(r)}{r}-\dfrac{zw'(r)}{r}+\epsilon_{m}.\label{eq:strain-components}
\end{equation}
In Eq.~\eqref{eq:strain-components}, the derivatives are considered with respect to the radial coordinate, $r$. The misfit strain in the system, $\epsilon_m$, is assumed to be accommodated in the film alone\cite{Freund1999}. Furthermore, the formulation assumes large rotations but small strains.

\subsection{\label{sec:level2}Curvature-mismatch relation for thin Si(001) wafer substrate}
 In the Si(001) wafer, the plane of the wafer is perpendicular to the [001] direction, which is along the $z$-axis of the deformation coordinate system. The $r$ and $\theta$ directions of the deformation coordinate axes are represented by two orthogonal directions in the plane of the single crystal wafer. Hence, the crystallographic axes of the Si(001) wafer also coincide with the axes of deformation. The stress and strain tensors are related through the elastic stiffness matrix as\cite{Janssen2008}
\begin{equation}
\left(\begin{array}{c}\sigma_{rr}\\\sigma_{\theta\theta}\\\sigma_{zz}\\\sigma_{\theta z}\\\sigma_{rz}\\\sigma_{r\theta}\end{array}\right)=
\begin{pmatrix}c_{11}&c_{12}&c_{12}&0&0&0\\c_{12}&c_{11}&c_{12}&0&0&0\\c_{12}&c_{12}&c_{11}&0&0&0\\0&0&0&c_{44}&0&0\\0&0&0&0&c_{44}&0\\0&0&0&0&0&c_{44}\end{pmatrix}
\left(\begin{array}{c}\epsilon_{rr}\\\epsilon_{\theta\theta}\\\epsilon_{zz}\\2\epsilon_{\theta z}\\2\epsilon_{rz}\\2\epsilon_{r\theta}\end{array}\right).\label{eq:stress-strain}
\end{equation}
Here $\sigma_{ij}$: components of stress tensor, $c_{ij}$: elastic stiffness constants of Si and
$\epsilon_{ij}$: components of strain tensor.\\
In the substrate material, from Eq.~\eqref{eq:stress-strain}
\begin{align}
\sigma_{zz}&=c_{11}\epsilon_{rr}+c_{12}\epsilon_{\theta\theta}+c_{11}\epsilon_{zz}=0,\nonumber \\
\implies\epsilon_{zz}&=-\dfrac{c_{12}}{c_{11}}(\epsilon_{rr}+\epsilon_{\theta\theta}), \nonumber \\
\implies\sigma_{rr}&=\left(\dfrac{c_{11}^2-c_{12}^2}{c_{11}}\right)\epsilon_{rr}+\left(\dfrac{c_{11}c_{12}-c_{12}^2}{c_{11}}\right)\epsilon_{\theta\theta},\label{eq:sigma-rr}\\
\implies\sigma_{\theta\theta}&=\left(\dfrac{c_{11}c_{12}-c_{12}^2}{c_{11}}\right)\epsilon_{rr}+\left(\dfrac{c_{11}^2-c_{12}^2}{c_{11}}\right)\epsilon_{\theta\theta}\label{eq:sigma-tt}.
\end{align}
From equations ~\eqref{eq:energy-density-expanded}, ~\eqref{eq:sigma-rr} and ~\eqref{eq:sigma-tt}, the elastic strain energy density in the substrate material can be written as
\begin{equation}
U^s(r,z)=\dfrac{1}{2}\left[\left(\dfrac{c_{11}^2-c_{12}^2}{c_{11}}\right)\left(\epsilon_{rr}^2+\epsilon_{\theta\theta}^2\right)+2\left(\dfrac{c_{11}c_{12}-c_{12}^2}{c_{11}}\right)\left(\epsilon_{rr}\epsilon_{\theta\theta}\right)\right].\label{eq:energy-density-subs}
\end{equation}
In order to preserve the uniform curvature ($\kappa$) assumption, the parametric forms for the substrate mid-plane deflections are adopted as\cite{Freund1999}
\begin{equation}
u(r)=\epsilon_0 r + \epsilon_1 r^3 ,\quad w(r)=\frac{\kappa r^2}{2}.\label{eq:displacements}
\end{equation}
Substituting Eq.~\eqref{eq:displacements} in Eq.~\eqref{eq:strain-components} results in
\begin{align}
\epsilon_{rr}&=\epsilon_0+3\epsilon_1r^2+\dfrac{\kappa^2r^2}{2}-\kappa z+\epsilon_m, \quad \epsilon_{\theta\theta}=\epsilon_0+\epsilon_1r^2-\kappa z+\epsilon_m,\nonumber\\
\implies \epsilon_{rr}&=\epsilon_{\theta\theta}+r^2\left[2\epsilon_1+\dfrac{\kappa^2}{2}\right]=\epsilon_{\theta\theta}+r^2\alpha.\label{eq:strain-curvature}
\end{align}
Using Eq.~\eqref{eq:strain-curvature}, Eq.~\eqref{eq:energy-density-subs} reduces to
\begin{align}
U^s(r,z)&=\dfrac{c_{11}^2+c_{11}c_{12}-2c_{12}^2}{c_{11}}\left(\epsilon_{\theta\theta}^2+\epsilon_{\theta\theta}r^2\alpha\right)+\dfrac{c_{11}^2-c_{12}^2}{c_{11}}\left(\dfrac{r^4\alpha^2}{2}\right), \label{eq:energy-density-subs-simple}
\end{align}
For the elastically isotropic film, the strain energy density, $U^f(r,z)$ for the plane stress case can be written as
\begin{align}
U^f(r,z)&=\dfrac{1}{2}\dfrac{E_f}{1-\nu_f^2}[\epsilon_{rr}^2+\epsilon_{\theta\theta}^2
 +2\nu_f \epsilon_{rr}\epsilon_{\theta\theta}],\label{eq:energy-density-film}
\end{align}
where $\nu_f$ and $E_f$ denote the Poisson's ratio and the Young's modulus of elasticity of the film material, respectively.
Substituting Eq.~\eqref{eq:strain-curvature} in Eq.~\eqref{eq:energy-density-film} results in
\begin{align}
U^f(r,z)&=\dfrac{E_f}{1-\nu_f}\left(\epsilon_{\theta\theta}^2+\epsilon_{\theta\theta}r^2\alpha\right)+\dfrac{E_f}{1-\nu_f^2}\left(\dfrac{r^4\alpha^2}{2}\right),\nonumber \\ 
&=M_f\left(\epsilon_{\theta\theta}^2+\epsilon_{\theta\theta}r^2\alpha\right)+\dfrac{M_f}{1+\nu_f}\left(\dfrac{r^4\alpha^2}{2}\right). \label{eq:energy-density-film-simple}
\end{align}
Here, $M_f$ represents the biaxial modulus of the film material. Comparing equations ~\eqref{eq:energy-density-film-simple} and ~\eqref{eq:energy-density-subs-simple},\\ $(c_{11}^2+c_{11}c_{12}-2c_{12}^2)/c_{11}$ represents an equivalent biaxial modulus for the Si(001) wafer material. A similar observation was made by \citet{Brantley1973}, \citet{Janssen2008} and \citet{Sharan2016} in their respective works on Si wafer substrate systems. Hence, Eq.~\eqref{eq:energy-density-subs-simple} can be written as
\begin{equation}
U^s(r,z)=M_s\left(\epsilon_{\theta\theta}^2+\epsilon_{\theta\theta}r^2\alpha\right)+\dfrac{c_{11}^2-c_{12}^2}{c_{11}}\left(\dfrac{r^4\alpha^2}{2}\right).\label{eq:energy-density-subs-Ms}
\end{equation}
The total potential energy of the system in terms of $\epsilon_0$, $\epsilon_1$ and $\kappa$ is given by\cite{Freund1999}
\begin{align}
V(\epsilon_0,\epsilon_1,\kappa)
&=2\pi\int^R_0\int^{h_s/2}_{-h_s/2}U^s(r,z)r drdz  + 2\pi\int^R_0\int^{h_f+h_s/2}_{h_s/2}U^f(r,z)r drdz,\label{eq:potential-energy}\\
&=V^s(\epsilon_0,\epsilon_1,\kappa)+V^f(\epsilon_0,\epsilon_1,\kappa). \nonumber
\end{align}
For the equilibrium condition  of stationary potential energy to hold, $\partial V / \partial \kappa$, $\partial V / \partial{\epsilon_0}$ and $\partial V / \partial{\epsilon_1}$ must be equal to zero.
The potential energy of the substrate material can be written as
\begin{equation}
V^s(\epsilon_0,\epsilon_1,\kappa)=2\pi\int^R_0\int^{h_s/2}_{-h_s/2}\left[M_s\left(\epsilon_{\theta\theta}^2+\epsilon_{\theta\theta}r^2\alpha\right)+\dfrac{c_{11}^2-c_{12}^2}{c_{11}}\left(\dfrac{r^4\alpha^2}{2}\right)\right]r drdz.\label{eq:potential-energy-subs}
\end{equation}
The potential energy of the film material can be written as
\begin{equation}
V^f(\epsilon_0,\epsilon_1,\kappa)=2\pi\int^R_0\int^{h_f+h_s/2}_{h_s/2}\left[M_f\left(\epsilon_{\theta\theta}^2+\epsilon_{\theta\theta}r^2\alpha\right)+\dfrac{M_f}{1+\nu_f}\left(\dfrac{r^4\alpha^2}{2}\right)\right]r drdz.\label{eq:potential-energy-film}
\end{equation}
Solving $\partial V / \partial{\epsilon_0}=0$ and $\partial V / \partial{\epsilon_1}=0$ for $\epsilon_0$ and $\epsilon_1$ in terms of $\kappa$, and substituting them back in $\partial V / \partial \kappa=0$ gives the desired expression for the curvature. These equations have been simplified for an analogous case when the substrate material is isotropic and $h_f/h_s\ll1$, by \citet{Freund2000}. Extending the same approach to the case when the substrate material is made from Si(001) wafer, the condition for stationary potential energy results in
\begin{equation}
S=K\left[1+\left(\dfrac{c_{11}}{c_{11}+c_{12}}\right)K^2\right],\label{eq:si001-largedef}
\end{equation}
where,
$S=\dfrac{3}{2}\epsilon_mR^2h_fM_f/(h_s^3Ms)$ and $K=\dfrac{1}{4}R^2\kappa/h_s$ are the normalized mismatch strain and normalized curvature, respectively.

\subsection{\label{sec:level2}Curvature-mismatch relation for thin Si(111) wafer substrate}

In the Si(111) wafer, the plane of the wafer is perpendicular to the [111] direction. The constitutive equation is written in the frame of the Si crystal while the deformation of the substrate takes place in the frame of the wafer, which in this case do not match. The reference frame of the constitutive equation in Si(111) wafer has been transformed to align with the coordinate system describing the deformation of the substrate. Such transformation allows for precise representation of the deformation field. The components of the compliance matrix in the transformed frame can be obtained from \citet{Janssen2008}. Thus, the radial and circumferential strains can be written as
\begin{align}
\tilde{\epsilon}_{rr}&=\left(\dfrac{s_{11}}{2}+\dfrac{s_{12}}{2}+\dfrac{s_{44}}{4}\right)\tilde{\sigma}_{rr}+\left(\dfrac{s_{11}}{6}+\dfrac{5s_{12}}{6}-\dfrac{s_{44}}{12}\right)\tilde{\sigma}_{\theta\theta} \notag \\
&=A\tilde{\sigma}_{rr}+B\tilde{\sigma}_{\theta\theta},\label{eq:epsilon-tilde-rr}\\
\tilde{\epsilon}_{\theta\theta}&=\left(\dfrac{s_{11}}{6}+\dfrac{5s_{12}}{6}-\dfrac{s_{44}}{12}\right)\tilde{\sigma}_{rr}+\left(\dfrac{s_{11}}{2}+\dfrac{s_{12}}{2}+\dfrac{s_{44}}{4}\right)\tilde{\sigma}_{\theta\theta} \notag \\ &=B\tilde{\sigma}_{rr}+A\tilde{\sigma}_{\theta\theta}.\label{eq:epsilon-tilde-tt}
\end{align}
Rewriting equations ~\eqref{eq:epsilon-tilde-rr} and ~\eqref{eq:epsilon-tilde-tt} for $\tilde{\sigma}_{rr}$ and $\tilde{\sigma}_{\theta\theta}$ 
\begin{align}
\tilde{\sigma}_{rr}&=\dfrac{1}{A^2-B^2}(A\tilde{\epsilon}_{rr}-B\tilde{\epsilon}_{\theta\theta}),\label{eq:sigma-tilde-tt}\\
\tilde{\sigma}_{\theta\theta}&=\dfrac{1}{A^2-B^2}(A\tilde{\epsilon}_{\theta\theta}-B\tilde{\epsilon}_{rr}).\label{eq:sigma-tilde-rr}
\end{align}
From equations ~\eqref{eq:energy-density-expanded} and ~\eqref{eq:strain-curvature}
\begin{align}
U^s(r,z)&=\dfrac{1}{A+B}\left(\epsilon_{\theta\theta}^2+\epsilon_{\theta\theta}r^2\alpha\right)+\dfrac{A}{A^2-B^2}\left(\dfrac{r^4\alpha^2}{2}\right),\notag \\
&=\dfrac{6}{4s_{11}+8s_{12}+s_{44}}\left(\epsilon_{\theta\theta}^2+\epsilon_{\theta\theta}r^2\alpha\right)+\dfrac{A}{A^2-B^2}\left(\dfrac{r^4\alpha^2}{2}\right),\notag \\
&=M_s'\left(\epsilon_{\theta\theta}^2+\epsilon_{\theta\theta}r^2\alpha\right)+\dfrac{A}{A^2-B^2}\left(\dfrac{r^4\alpha^2}{2}\right).\label{eq:energy-densiti-si111-subs}
\end{align}
Here, $M_s'$ represents the equivalent biaxial modulus of the Si(111) wafer material. Following the similar approach as with Si(001) wafer substrate, the curvature-mismatch relation for this case results in
\begin{equation}
S=K\left[1+\dfrac{4}{3}\left(\dfrac{2s_{11}+4s_{12}+\dfrac{1}{2}s_{44}}{2s_{11}+2s_{12}+s_{44}}\right)K^2\right].\label{eq:si111-largedef}
\end{equation}
In equations ~\eqref{eq:si001-largedef} and ~\eqref{eq:si111-largedef}, the film may be anisotropic or isotropic. For an anisotropic film, an appropriate biaxial modulus may be used in the place of $M_f$ to calculate the normalized mismatch strain $S$.

\section{Numerical Results and Discussion}
\begin{figure}[!h]
\begin{center}
\subfloat[]
{
\includegraphics[width=0.52\textwidth]{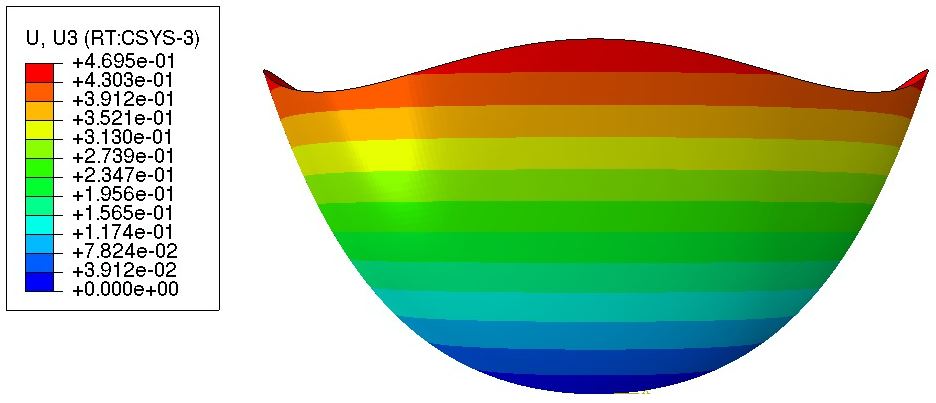}
}
\subfloat[]
{
\includegraphics[width=0.52\textwidth]{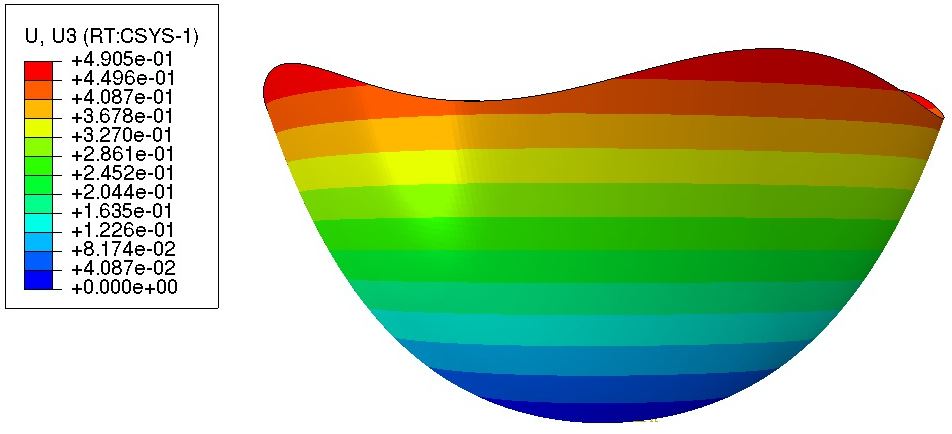}
}
\end{center}
\caption{Deflection along z-axis induced in (a) Si(001) substrate and (b) Si(111) substrate, in $mm$.}
\label{fig:Figure-2}
\end{figure}

The deformations are studied with simulations performed using commercial finite element software Abaqus\cite{Abaqus2013}. The thin-film configuration is modeled using four-noded composite shell elements. This choice allows a distribution in material properties across the thickness of the shell. The geometric non-linearities due to large rotations (but small strains) are also accounted for in the simulation. In Fig.~\ref{fig:Figure-1}, the parameters $h_s/h_f$ and $R/h_s$ are fixed at 20 and 50, respectively. Following this, the undeformed radius of the system is chosen to be 10 mm. The mismatch strain is provided to the system in the form of thermal mismatch. For the ease of numerical simulation, the film and substrate materials in both cases have been chosen to have identical mechanical properties, but varying thermal expansion coefficients. In Fig.~\ref{fig:Figure-2}, the thermal expansion coefficient of the bottom layer is taken to be $10^{-5}$/ \textdegree C and that of the top layer is set at $0$/ \textdegree C, while both layers are subjected to the same temperature rise to produce the appropriate mismatch strain (note that the mismatch strain ultimately depends on the difference and not the individual thermal expansion coefficients of each layer). Also, the ratio of biaxial moduli ($M_f/M_s$) is one in each simulation. The stiffness constants of silicon obtained by \citet{McSkimin1964} are used in order to simplify equations ~\eqref{eq:si001-largedef} and ~\eqref{eq:si111-largedef}. The biaxial moduli used for Si(001) and Si(111) layers in the simulation are $1.803$ x $10^{-11} Nm^{-2}$ and $2.291$ x $10^{-11} Nm^{-2}$, respectively.\cite{Janssen2008}
The transverse deflections for the Si(001) substrate along the [1$\overline{1}$0], [110], [$\overline{1}$10] and [$\bar{1}$$\bar{1}$0] directions are computed for varying normalized mismatch strains. A similar approach was used by \citet{Janssen2008} in order to determine the average curvature of anisotropic substrates for small deformations, experimentally. The radial curvature along each direction, $k(r)$ is calculated by first fitting an eighth order polynomial in $r$, $w(r)$ to the deflection data and then determining the curvature as $w^{''}(r)/{(w'(r)^2+1)}^{3/2}$. The normalized curvature along each direction as a function of $r$ is calculated as $K(r)=R^2k(r)/4h_s$. The function value at each radial position is then averaged over the four directions. A similar approach is followed to calculate the average normalized curvature as a function of $r$ for the Si(111) substrate configuration. Here, the deflections along the $z$-axis are measured along the [01$\overline1$], [1 1 $\overline{2}$], [10$\overline{1}$], [2 $\overline{1}$ $\overline{{1}}$], [1$\overline1$0] and [1 $\overline{2}$ 1] directions, each separated by an angle of 30 degrees.
\begin{figure}[!h]
\begin{center}
\subfloat[]
{
\includegraphics[width=0.48\textwidth]{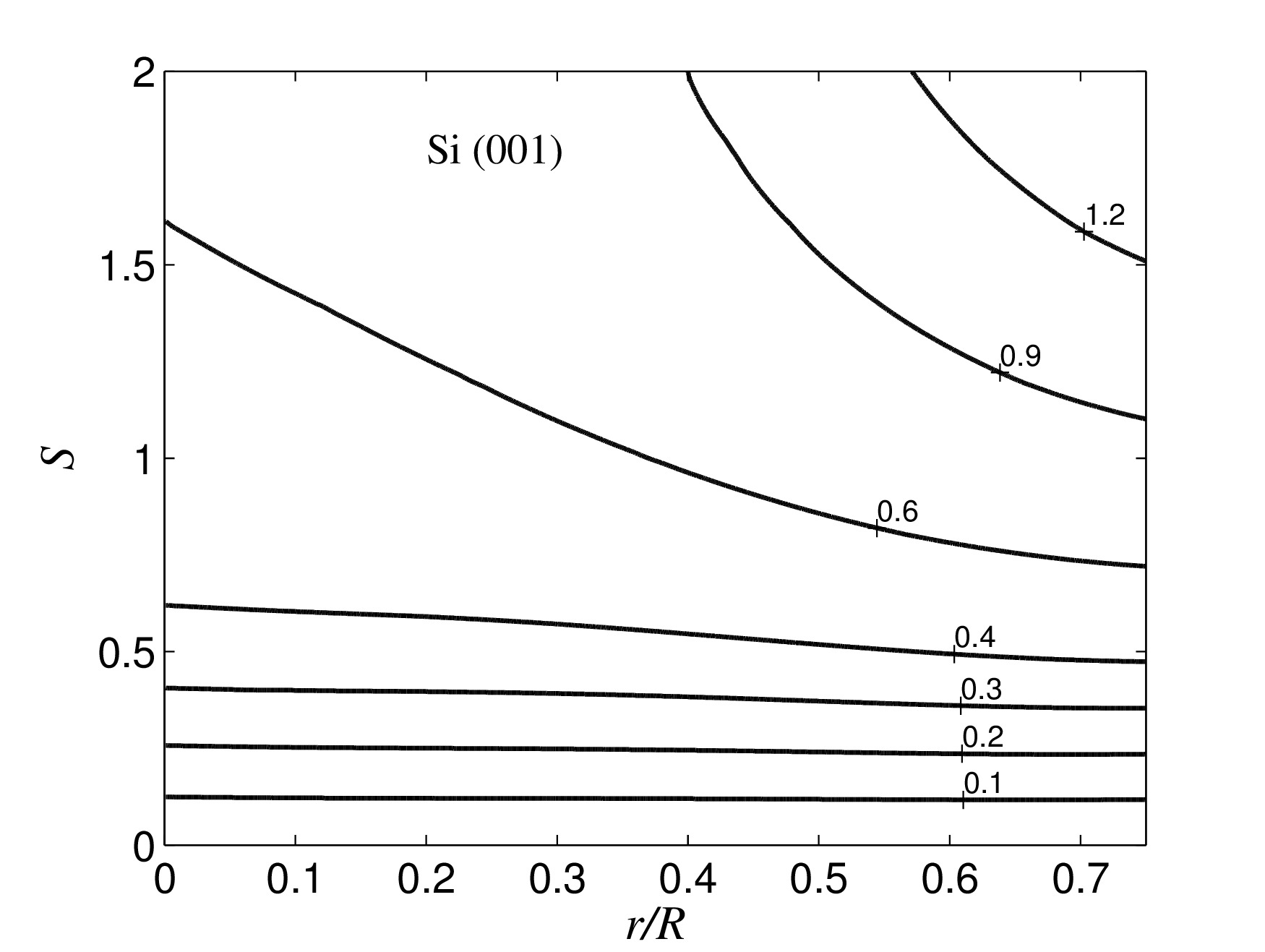}
}
\subfloat[]
{
\includegraphics[width=0.48\textwidth]{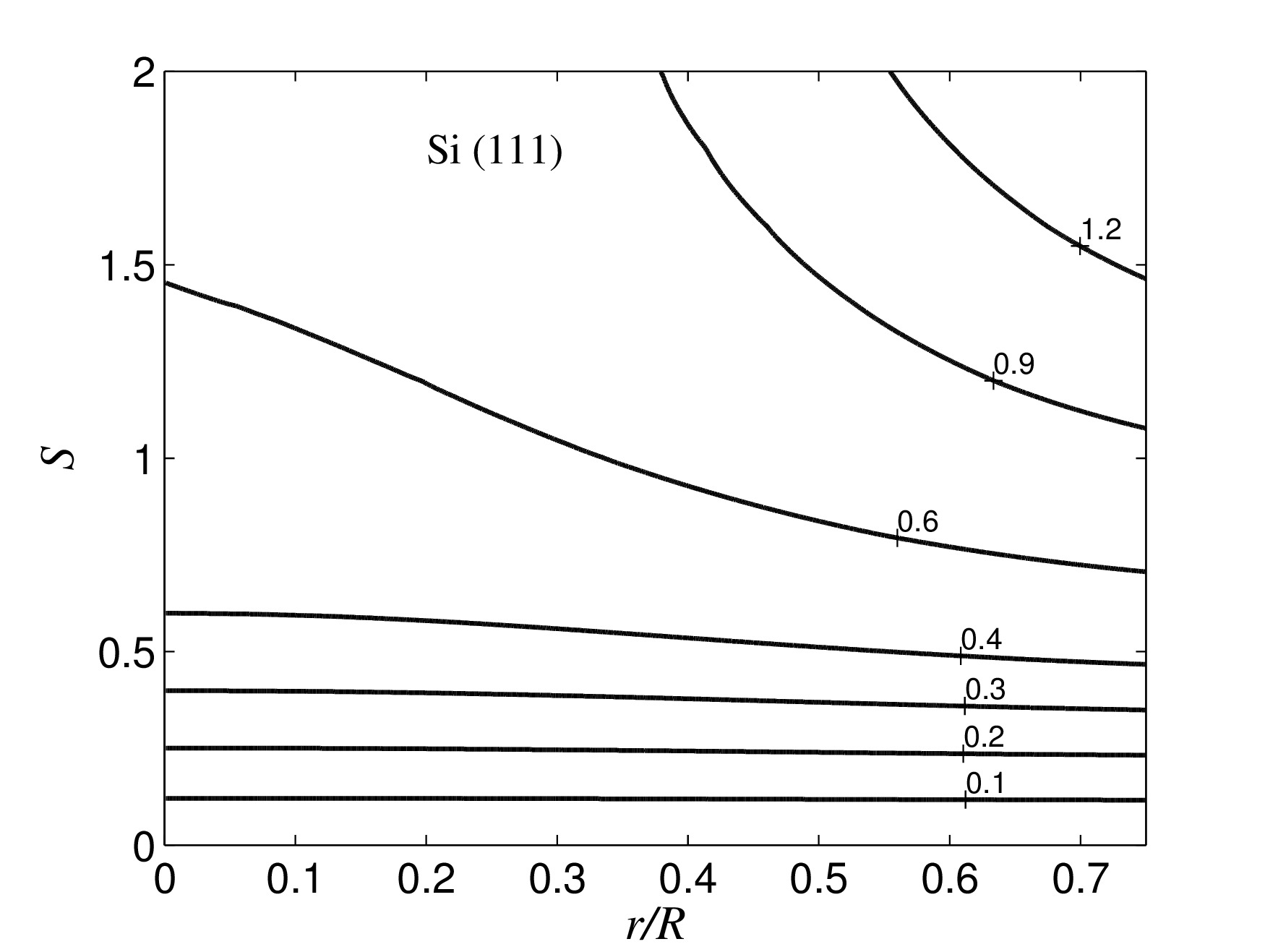}
}
\end{center}
\caption{Contour plots of normalized curvature $K$ as the normalized mismatch strain $S$ and normalized distance $r/R$ are varied for the system with (a) Si(001) substrate and (b) Si(111) substrate.}
\label{fig:Figure-3}
\end{figure}\begin{figure}[!h]
\begin{center}
\subfloat[]
{
\includegraphics[width=0.48\textwidth]{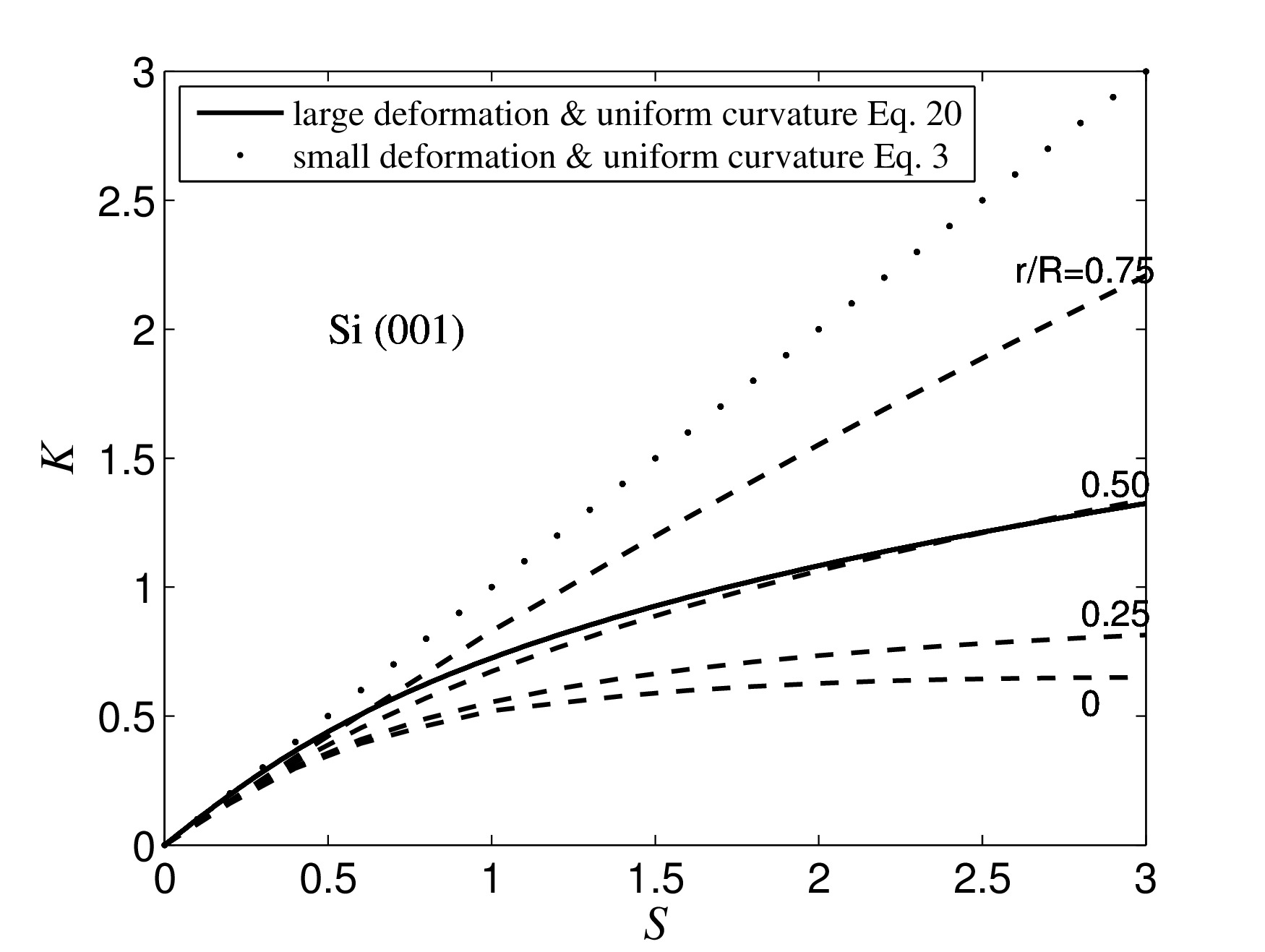}
}
\subfloat[]
{
\includegraphics[width=0.48\textwidth]{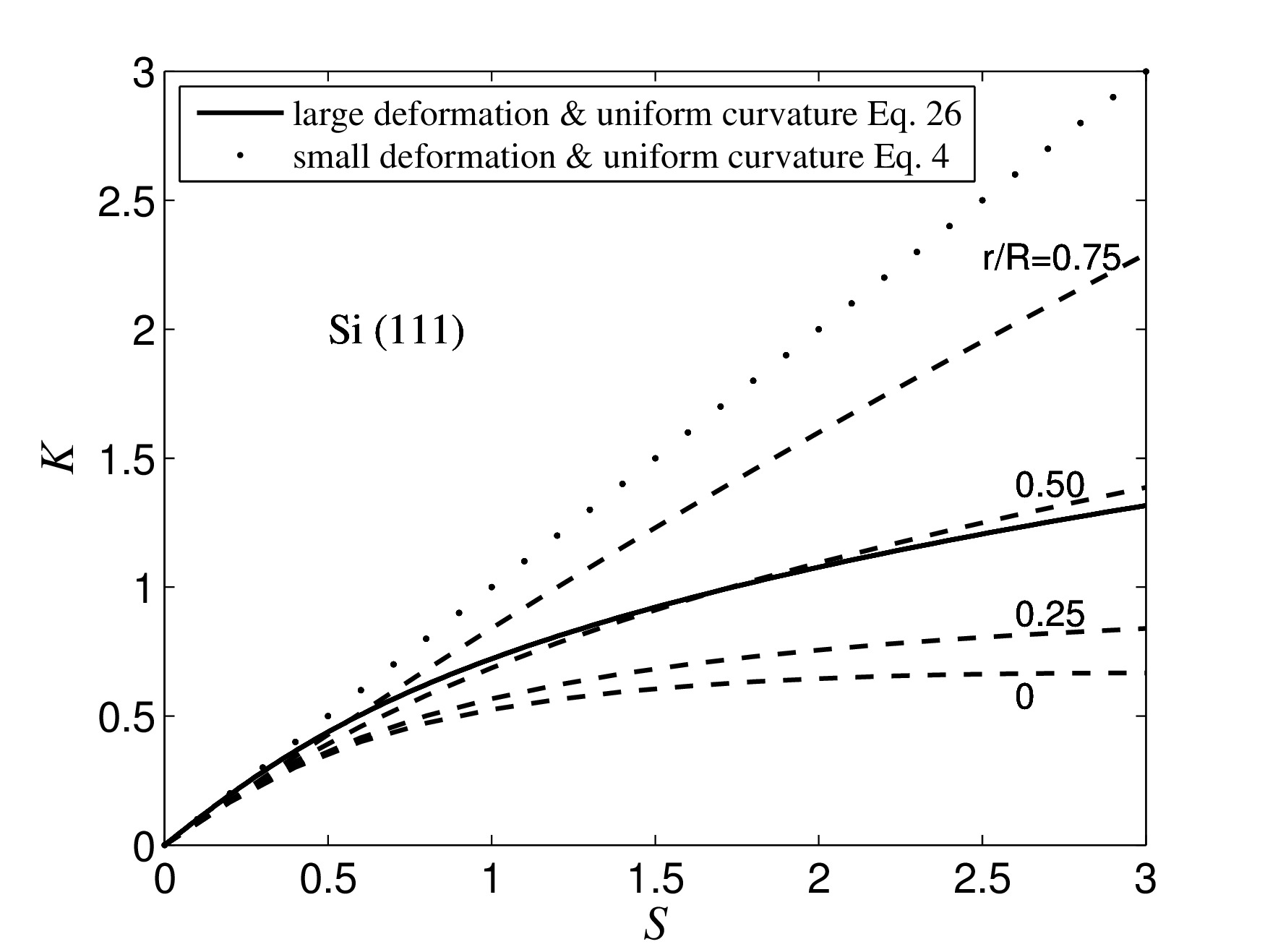}
}
\end{center}
\caption{Plot for normalized curvature $K(r)$ vs the normalized mismatch strain $S$ for small and large deformations for the system with (a) Si(001) substrate and (b) Si(111) substrate. The dashed lines plot $K$ obtained from the finite element simulation at four values of $r/R$ starting from $r/R=0$.}
\label{fig:Figure-4}
\end{figure}\begin{figure}[!h]
\begin{center}
\subfloat[]
{
\includegraphics[width=0.48\textwidth]{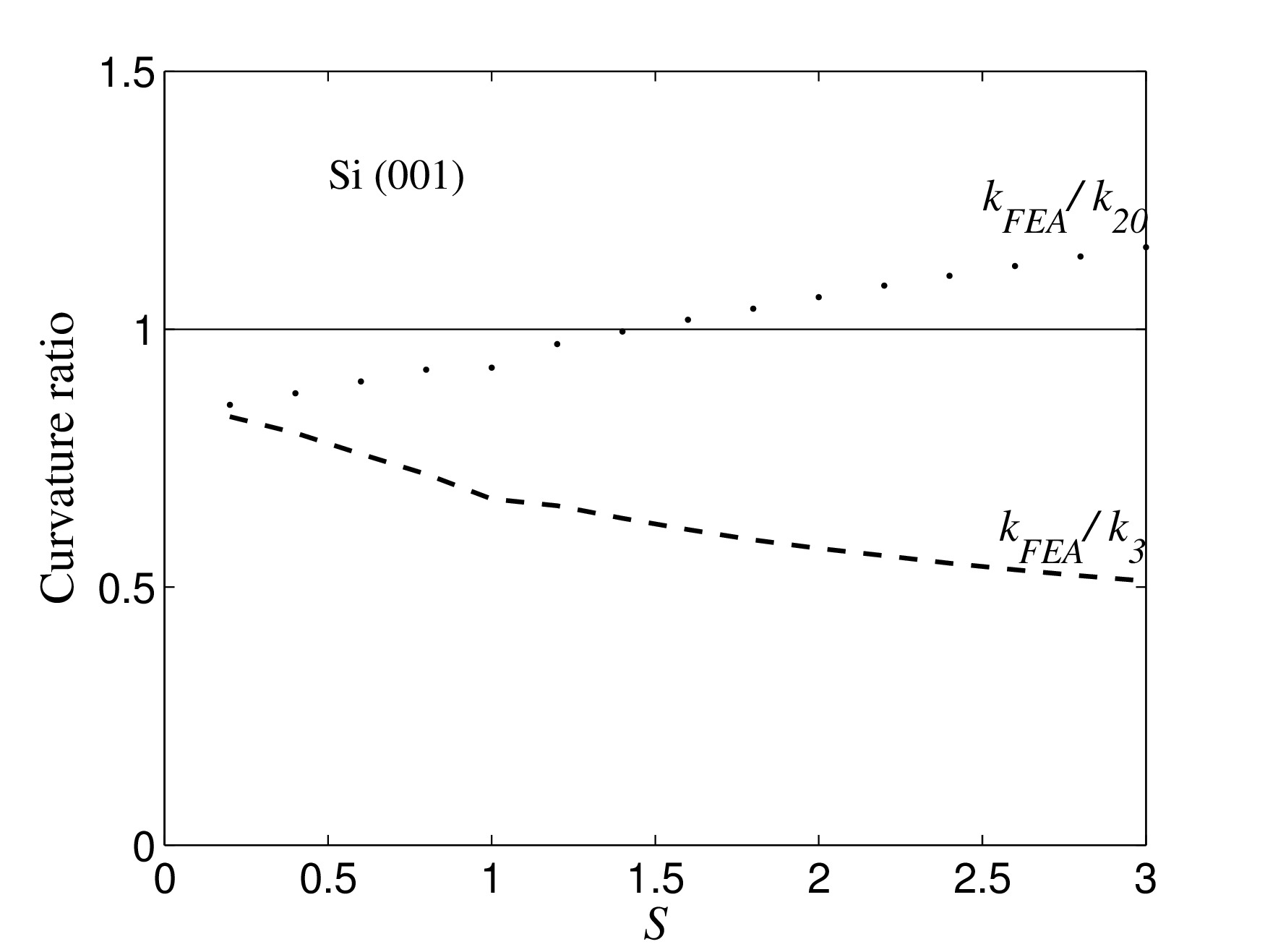}
}
\subfloat[]
{
\includegraphics[width=0.48\textwidth]{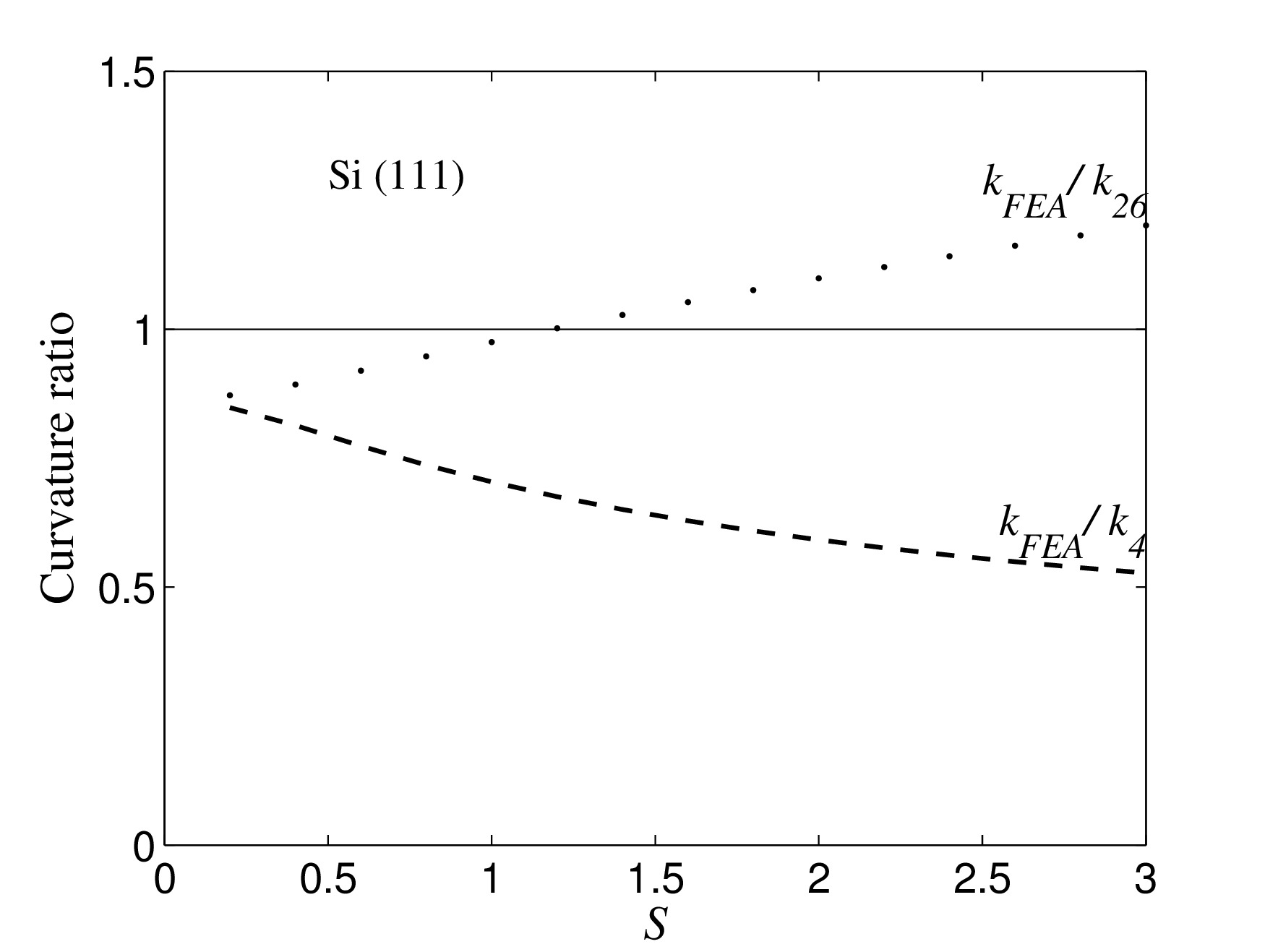}
}
\end{center}
\caption{Deviation of the curvatures obtained from (a) equations ~\eqref{eq:si001-largedef} and ~\eqref{eq:stoney-si001}, and (b) equations ~\eqref{eq:si111-largedef} and ~\eqref{eq:stoney-si111}, from the finite element simulation, plotted with respect to $S$.}
\label{fig:Figure-5}
\end{figure}
Fig.~\ref{fig:Figure-3} shows the plots between normalized mismatch ($S$) and normalized distance ($r/R$) for different normalized curvature ($K$) values. The figure indicates uniform curvature across the substrate mid plane for smaller values of $S$, for both Si(001) and Si(111) substrates. As the mismatch strain is increased, the curvature is increasingly non-linear. Hence, the uniform curvature assumption nearly holds good for normalized mismatch strains that are less than 0.3. A similar observation was made by \citet{Freund2000} for thin film configurations with isotropic substrates undergoing large deformations. On substituting isotropic material properties for the substrate in the numerical simulation, the results obtained by \citet{Freund2000} have been reproduced (results not shown here). Furthermore, the non-uniform curvature occurs at a smaller mismatch for anisotropic substrates when compared to the isotropic substrate used in Freund's work \cite{Freund2000}. This observation is possibly due to the direction dependence of curvature in anisotropic substrates. Fig.~\ref{fig:Figure-4} shows the plots between normalized curvature ($K$) and normalized mismatch strain ($S$) for different $r/R$ values. Also the curvature obtained from the present large deformation analysis (solid line) is compared with small deformation (dotted line) and finite element results (dashed lines). The figure indicates that the curvatures obtained from equations ~\eqref{eq:si001-largedef} and ~\eqref{eq:si111-largedef} (i.e., large deformation equations) lie within the curvature values obtained from the finite element simulations. Whereas, equations ~\eqref{eq:stoney-si001} and ~\eqref{eq:stoney-si111} (i.e., small deformation equations) overestimate the curvature. In Fig.~\ref{fig:Figure-5}, $k_{~\eqref{eq:si001-largedef}}$ and $k_{~\eqref{eq:si111-largedef}}$ correspond to curvatures obtained from equations ~\eqref{eq:si001-largedef} and ~\eqref{eq:si111-largedef}, respectively. $k_{~\eqref{eq:stoney-si001}}$ and $k_{~\eqref{eq:stoney-si111}}$ are curvatures obtained from equations ~\eqref{eq:stoney-si001} and ~\eqref{eq:stoney-si111}, respectively. In figures~\ref{fig:Figure-5}(a) and~\ref{fig:Figure-5}(b), $k_{FEA}$ is the curvature estimated from the finite element simulations for Si(001) and Si(111) wafer substrates, respectively. $k_{FEA}$ is calculated by taking the average of curvatures for a given mismatch strain, over the four equally spaced radii starting from the substrate centre. It can be observed that the deviations are much larger for curvatures obtained from equations ~\eqref{eq:stoney-si001} and ~\eqref{eq:stoney-si111} when compared to equations ~\eqref{eq:si001-largedef} and ~\eqref{eq:si111-largedef}, indicating that the derived formulae are a better fit to the data obtained from the finite element simulations, as they account for large deformations in the configuration. Furthermore, a semi-analytical model for the curvature (at $r$=0) of Si-doped GaN on Si (111) substrate undergoing large deformations, was presented by \citet{Clos2004}. The deviation of this central curvature with respect to the curvature obtained from Eq.~\eqref{eq:si111-largedef} is about 32\% for a thickness ratio ($h_f/h_s$) of 0.01. Whereas, this deviation is as high as 83\% for the same thickness ratio, when the curvature is calculated from the small deformation Stoney formula (Eq.~\eqref{eq:stoney-si111}). This further supports the use of Eq.~\eqref{eq:si111-largedef} over Eq.~\eqref{eq:stoney-si111} for Si (111) substrates deforming in the non-linear regime.
\section{Conclusions}
Normalized curvature-mismatch relations are derived for thin films bonded to anisotropic substrates (equations ~\eqref{eq:si001-largedef} and ~\eqref{eq:si111-largedef}) undergoing large deformations. The equilibrium condition of stationary potential energy is used in order to arrive at these equations. Equations ~\eqref{eq:si001-largedef} and ~\eqref{eq:si111-largedef} can be used also for elastically anisotropic films by substituting an appropriate biaxial modulus for $M_f$, depending on the material. The formulae derived for large deformations along with existing expressions for small deformations for anisotropic substrates are compared with numerical results obtained using Abaqus FEA. Furthermore, the curvature obtained from the numerical data is almost uniform across the radius of the substrate for normalized mismatch ($S$) values within 0.3, for both Si(001) and Si(111) substrates. The direction dependence of curvature in anisotropic substrates is evident from the observation that non-uniformity in substrate curvature occurs at a smaller mismatch strain when compared to isotropic substrates. The analytical formulae derived for large deformations (equations ~\eqref{eq:si001-largedef} and ~\eqref{eq:si111-largedef}) match the finite element results better, when compared to the formulae derived by \citet{Janssen2008}, which were derived in the small deformation regime.

\end{document}